\pdfoutput=1 
\documentclass[
superscriptaddress,
 aps,
reprint
]{revtex4-1}

\usepackage{graphicx}
\usepackage{dcolumn}
\usepackage{bm}
\usepackage{amsmath}
\usepackage{color}
\usepackage{float}
\usepackage{array,url,kantlipsum}
\usepackage{verbatim}
\usepackage{xcolor}
\usepackage{CJKutf8}
\usepackage{amssymb}
\usepackage[colorlinks=true, linkcolor=blue, citecolor=blue, urlcolor=blue]{hyperref}
\usepackage{soul}
\usepackage{cuted}
\usepackage[mathlines]{lineno}
\usepackage{amsmath,lipsum}
\makeatletter

\newcommand{\Rmnum}[1]{\expandafter\@slowromancap\romannumeral #1@}
\makeatother
\newcolumntype{P}[1]{>{\centering\arraybackslash}p{#1}}

\begin{document}
\preprint{APS/123-QED}

\title{Detuning-insensitive wide-field imaging of vector microwave fields with diamond sensors}
\author{Xiu-Qi Chen}
\author{Rui-Zhi Zhang}
\affiliation{Beijing National Laboratory for Condensed Matter Physics, Institute of Physics, Chinese Academy of Sciences, Beijing 100190, China}
\affiliation{School of Physical Sciences, University of Chinese Academy of Sciences, Beijing 100049, China}

\author{Gang-Qin Liu}
\affiliation{Beijing National Laboratory for Condensed Matter Physics, Institute of Physics, Chinese Academy of Sciences, Beijing 100190, China}

\author{Huijie Zheng}
\email{hjzheng@iphy.ac.cn}
\affiliation{Beijing National Laboratory for Condensed Matter Physics, Institute of Physics, Chinese Academy of Sciences, Beijing 100190, China}

 \date{\today}

\begin{abstract}
Nitrogen vacancy (NV) centers in diamond have precipitated profound advances in microwave detection, manifesting themselves both in spatial resolution and sensitivity.
However, typical methods based on Rabi oscillations are subject to detunings due to thermal and magnetic fluctuations and/or gradients, which introduce systematic errors and render the measurements susceptible to environmental perturbations. Here, we propose and demonstrate a novel approach for determining both the magnitude and direction of microwaves, by exploiting the spectral line broadening effect in the optically detected magnetic resonance of NV centers. 
This method eliminates the requirement of aligning the MW frequency to the spin transitions and is therefore immune to variations and inhomogeneities of the magnetic field and temperature, providing an optimal tool for fast imaging applications. With this method, we achieved wide-field imaging of near field microwaves generated with a microscale $\rm{\Omega}$-pattern antenna with a resolution of 800\,nm. Combining with the vector detection using multi-axis NVs, a full reconstruction of the vector microwave fields is obtained.
Besides, our scheme also exhibits excellent linearity over a broad range of MW amplitudes, and the scale is theoretically calculated to be more than four orders. Our results augment the applicability of diamond-based microwave devices in applications under complex scenarios, especially where large dynamic range, fast test speed, and high spatial resolution are demanded.

\end{abstract}

\pacs{Valid PACS appear here}
\maketitle




\section{Introduction}


The detection of microwaves (MWs) is at the core of numerous modern technologies, including communication engineering\,\cite{holl2017holography}, biomedical diagnosis\,\cite{khalid2024emerging,yago2023effective,wang2023microwave}, semiconductor device design\,\cite{berweger2015microwave,huber2012calibrated}, quantum information processing\,\cite{you2011atomic,devoret2013superconducting}, non-destructive testing\,\cite{abou2023detection,li2017microwaves}, and radio-frequency device design\,\cite{rable2025flux,fan2024optimization}. While it is of key importance to improve the sensitivity in most applications, vector imaging MWs with high temporal and spatial resolution is also crucial. For instance, visualizing the electromagnetic near-field distribution is essential for identifying multipath interference or hotspot regions, thereby mitigating signal distortion and enhancing design robustness in advanced integrated circuit industry\,\cite{kosen2024signal}.
Despite a plethora of sensing protocols demonstrating exceptional sensitivity for weak MWs, fast imaging of the near-field distribution of vector MW fields, especially for strong MWs, remains challenging. 
The reason lies in the unavoidable crosstalk and/or side effects induced by the microwave radiation, e.g. thermal fluctuations, broadening effects for quantum sensors, etc, which would deteriorate the sensitivity of measurements. 


The nitrogen-vacancy (NV) center in diamond has been considered as one of the most promising candidates for MWs sensing\,\cite{wang2015high,basso2025wide,wang2022picotesla,li2023near,lamba2024vector,fan2024optimization,yang2018noninvasive,basso2025wide}, due to its exceptional performance in sensitivity, bandwidth, spatial and temporal resolution, chemical stability, and wide operational temperature range\,\cite{taylor2008high,zhou2014quantum,fan2024quantum}. Numerous proposals have been demonstrated to leverage NV centers to detect weak MW fields, with ongoing efforts focused on improving the sensitivity. 
For example, 
Wang\,\textit{ et al} proposed a continuous heterodyne detection scheme that improved the sensitivity to picotesla level\,\cite{wang2022picotesla}. Basso\,\textit{ et al} demonstrated a wide-field imaging of weak MWs on the Rabi oscillation frequency of NV center and the sensitivity is enhanced by a differential detection\,\cite{basso2025wide}. Besides Rabi oscillations, MWs can alternatively be estimated via the contrast of continuous-wave optical detection magnetic resonance (CW-ODMR) of the NV center \,\cite{fan2024optimization,jia2024near}.

However, in the case of strong microwave fields, some of the aforementioned methods suffer from the side effects, and become insufficient and inaccurate in measurement. For instance, methods based on Rabi oscillations are susceptible to detunings between the MW frequency to the spin transitions, which are typically induced by thermal and magnetic variations. These detunings introduce systematic errors and necessitate stringent environmental control for accurate measurements\,\cite{wang2015high}. Moreover, for wide-field imaging, it is necessary to calibrate the resonant frequency at each pixel before the Rabi measurement, which would otherwise lead to systematic errors, making it suboptimal in fast imaging applications. 

Methods based on ODMR contrast are considered as a non-quantized measurement, since, in addition to MW powers, the CW-ODMR contrast subjects to quite a few parameters, such as laser power, spin lifetime and also the coherence\,\cite{2011Avoiding}. Moreover, the contrast saturates at moderately high MW powers and thus loses sensitivity\,\cite{wang2022picotesla}. In contrast, the linewidth of ODMR transition increases due to an effect of power broadening and therefore can be used as an indicator for MW power in strong MW regime. 

In this work, we propose a protocol that leverages the power broadening effect for MW imaging that is immune to frequency detunings and overcomes the aforementioned limitations. The sensing regime is to modulate the amplitude of the microwaves (the target MWs) with a square wave (the modulation source), and subtract the difference in the linewidth of spin transition features in ODMR via lock-in detection. We conducted a combined theoretical and experimental investigation on the relationship between the differential linewidth of the ODMR and the magnitude of MW and achieved a sensitivity of 600 $\rm{nT/\sqrt{Hz}}$. 
Using a home-built wide-field imaging setup, the spatial distribution of a $\rm{\Omega}$ antenna is mapping and, furthermore, a full vector reconstruction ability is demonstrated by interrogating four orientation NVs. Theoretical simulation further shows that the scheme is capable of detecting a broad range of MW powers, maintaining a linear response for four orders of magnitude. This enables for precise quantification of the MW field with a wide dynamical range. 

\section{The principle}

The principle of the MW sensing relies on the broadening of the optically detected magnetic resonance (ODMR) spectrum, which is induced by the interaction between the NV center electron spin and the MW field. However, the linewidth of the ODMR spectrum can be influenced by numerous factors, for instance, laser power, magnetic field gradient\,\cite{mrozek2015longitudinal}, and electric field\,\cite{yu2024optically}. Thus, the absolute linewidth cannot be directly used to denote the MW strength. To mitigate systematic errors and enhance the sensitivity, we propose a protocol that modulates the amplitude of the MW signal with an auxiliary square wave and detects the difference in linewidth. We note that it is not necessary to modulate the target signal for lock-in detection as presented in this work, a similar job can be achieved by mixing it with a modulated auxiliary microwave and using heterodyne detection.

\begin{figure}[H]
    \centering
    \includegraphics[width=1\linewidth]{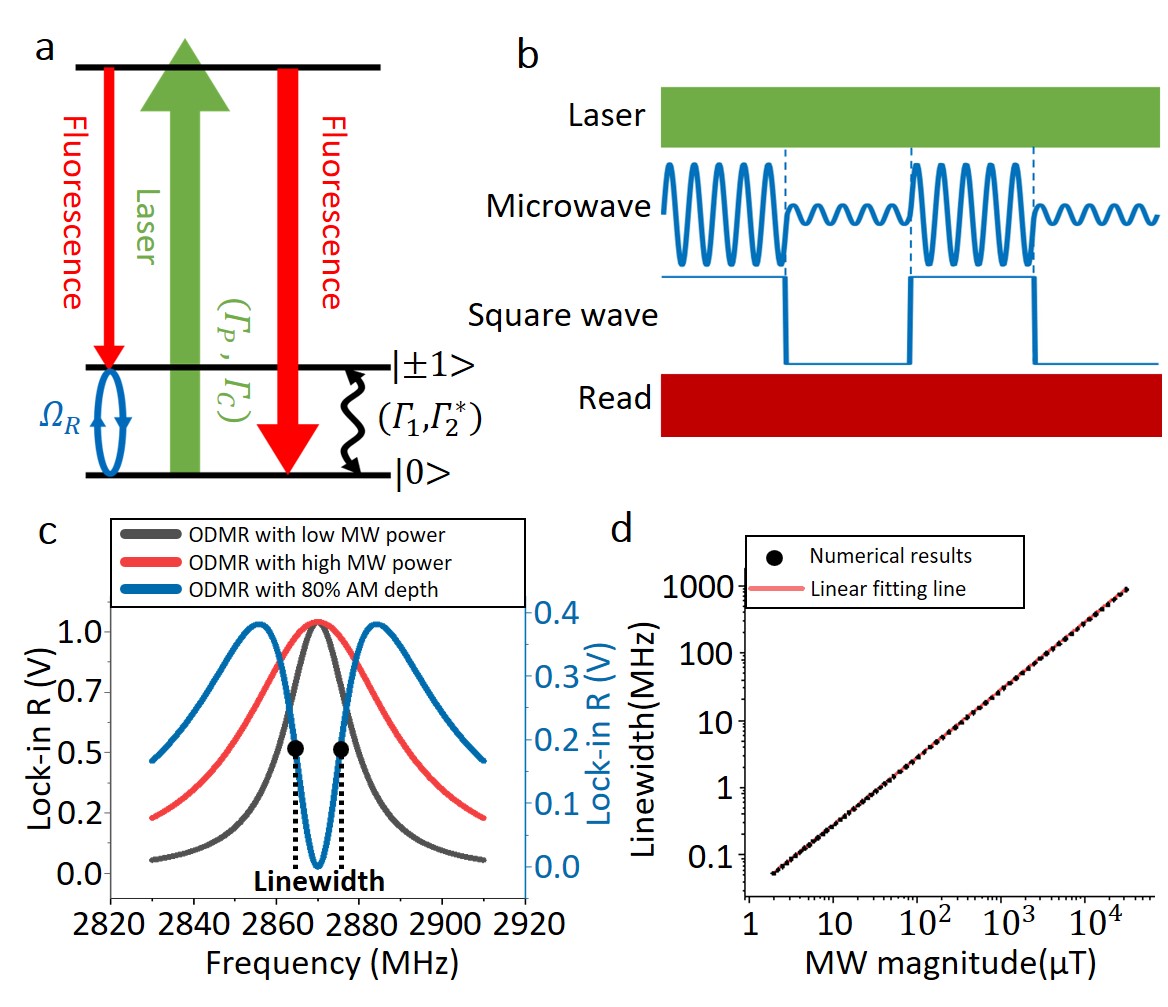}
    \caption{(a) The energy levels of the electron spin of the NV center, and their interaction with the MW field. (b) The sensing scheme of MW detection. The laser remains continuously activated while the MW signal is square-wave amplitude-modulated with a depth of 80\,\%. (c) The black curve represents the ODMR spectrum obtained under lower MW power, while the red corresponds to full MW power. The subtracted signal is shown by the blue curve.      
    (d) The theoretical dependence of the linewidth of the subtracted spectrum on the MW magnitude.}
    \label{fig:Principle}
\end{figure}

For simplicity, we use a two-level model to simulate the CW-ODMR: a 532-nm green laser continuously pumps the NV center into the state $|0\rangle$, and a continuous MW irradiation induces Rabi oscillations between the states $|0\rangle$ and $|1\rangle$, as depicted in Fig.\,\ref{fig:Principle}\,a. Here, $\Gamma_{1}$ and $\Gamma_{2}^{\ast}$ denote the spin relaxation rate and the inhomogeneous spin dephasing rate, respectively, $\Gamma_{p}$ and $\Gamma_{c}$ represent the optical polarization rate and the coherence relaxation rate due to optical pumping. Given the assumption that $\Gamma_{p}\gg\Gamma_{1}$ and $\Gamma_{c}\gg\Gamma_{2}^{\ast}$, the linewidth of the ODMR can be induced by both the laser and the microwave power. The profile of the spectral feature can be expressed as follows\,\cite{2011Avoiding}:
\begin{equation}
\setlength\abovedisplayskip{3pt}
\setlength\belowdisplayskip{3pt}
    I\left(\omega\right)=1-\frac{C}{1+\frac{\left(\omega-\omega_{c}\right)^{2}}{\Delta\omega^{2}}}\label{eq1}
\end{equation}
Where $C\propto\frac{\Omega_{R}^{2}}{\Omega_{R}^{2}+\Gamma^{\infty}_{p}\Gamma^{\infty}_{c}\left(\frac{S}{1+S}\right)^{2}}$ is the contrast of ODMR spectrum, $\Delta\omega=\frac{\Gamma_{c}^{\infty}}{2\pi}\sqrt{\left(\frac{S}{1+S}\right)^{2}+\frac{\Omega_{R}^{2}}{\Gamma_{c}^{\infty}\Gamma_{p}^{\infty}}}$ is the full width at half-maximum (FWHM) of the ODMR spectral line. \textit{S} is defined as the ratio of the effective power of the laser to its saturated power, \textit{$\Gamma_{p}^{\infty}$} the spin polarization rate at optical saturation, $\Gamma_{c}^{\infty}$ the rate of optical cycles at saturation, \textit{$\Omega_{R}$} the Rabi oscillation frequency without detuning and $\omega_{c}$ the zero-field ground-state splitting of NV center.
We defined $ I_{m}=\Omega_{R}^{2}$, which represents the power of MW. When the MW power is relatively high, and the rate at which the microwave induces electronic spin flips of the NV center is significantly faster than the laser polarization rate, i.e., when ${I_{m}\gg\Gamma^{\infty}_{p}\Gamma^{\infty}_{c}\left(\frac{S}{1+S}\right)^{2}}$ as illustrated in Eq.\,\eqref{eq1}, further increases in MW power will not lead to a larger contrast of the ODMR spectrum. In this scheme, the contrast loses sensitivity to MW power. Whereas the ODMR feature gets further broadened and the linewidth continues to increase complementarily. 
Considering the case of 80\,\% AM where MWs of intensities $I_{m}$ and $5 I_{m}$ are being alternately applied to the diamond, the ODMR features in different linewidths. 
Figure\,\ref{fig:Principle}\,b illustrates the sensing scheme of laser and MW. The black and red curves in Fig.\,\ref{fig:Principle}\,c represent the original ODMR spectra under conditions $ I_{m}$ and $5\,I_{m}$, respectively. The difference subtracted is then detected via a LIA, presented by the blue curve in Fig.\,\ref{fig:Principle}\,c. 
With such demodulation and subtraction, the contribution of MW powers to the linewidth is identified and isolated, thus improving the sensitivity and precision for MW detection. 
The derivation is illustrated as follows:

\begin{equation}
\setlength\abovedisplayskip{3pt}
\setlength\belowdisplayskip{3pt}
    L\left(\omega\right)\propto\frac{5\gamma I_{m}}{\gamma\left(5I_{m}+I_{s}\right)+\left(\omega-\omega_{c}\right)^{2}}-\frac{\gamma I_{m}}{\gamma\left(I_{m}+I_{s}\right)+\left(\omega-\omega_{c}\right)^{2}}\label{eq2}
\end{equation}
where $ I_{S}=\Gamma^{\infty}_{p}\Gamma^{\infty}_{c}\left(\frac{S}{1+S}\right)^{2}$ is defined as the saturating MW power density, $ \delta_{l}^{2}=\frac{\left(\Gamma^{\infty}_{c}\right)^{2}}{4\pi^{2}}\left(\frac{S}{1+S}\right)^{2}$ the laser power broadening factor, and $ \gamma=\delta_{l}^{2}/I_{S}$ the MW power broadening factor. During the testing process, the laser power was maintained at a constant level, ensuring that the $I_{S}$ remained unchanged.


To achieve the optimal sensitivity of the proposed scheme, the frequency difference at the two point of maximum slope of the subtracted ODMR dip (the two black dots illustrated in Fig.\,\ref{fig:Principle}c) is defined as the ``linewidth'' of the demodulated AM ODMR and used as an indicator to quantify the amplitude of the MW field. The frequency values for these two points are derived from the condition $ \frac{\partial^{2}L\left(\omega\right)}{\partial \omega^{2}}=0$.
By step-and-step derivation, the relationship between the linewidth of the AM ODMR and the applied MW magnitude is calculated and shown in Fig.\,\ref{fig:Principle}\,d. The values of\,$\gamma\cdot I_{s}$ can be estimated based on the intrinsic linewidth of the ODMR spectral line. Given samples containing ensemble NV centers with linewidth of 34\,kHz\,\cite{barry2024sensitive}, our scheme is capable to measure a field down to $\rm{\mu T}$.
The numerical simulation result shows that in strong MW field regime, when the MW power exceeds the saturation level, the linewidth of the AM ODMR spectrum still increases linearly with the MW amplitude within an amplitude ranging in four orders, and this capacity can be further increased by tuning the depth of modulation or alternatively the strength of the auxiliary field in a heterodyne measurement. Furthermore, a smaller $I_{s}$ is associated with a higher linewidth response to MW amplitude, indicating that the sensitivity of the MW detection scheme can be enhanced using lower laser power in the experimental setup (for details, see the section\,\Rmnum{4} in the supplementary materials). 
With this scheme, it does not require stringent alignment of the MW frequency to the spin transitions of NVs,
and effectively mitigates the systematic errors due to detunings that come from thermal and magnetic fluctuations and/or gradients, making it suitable for practical measurements under complex environments and wide-field imaging. 
 

\section{the Experiment}

\begin{figure}[H]
    \centering
    \includegraphics[width=1\linewidth]{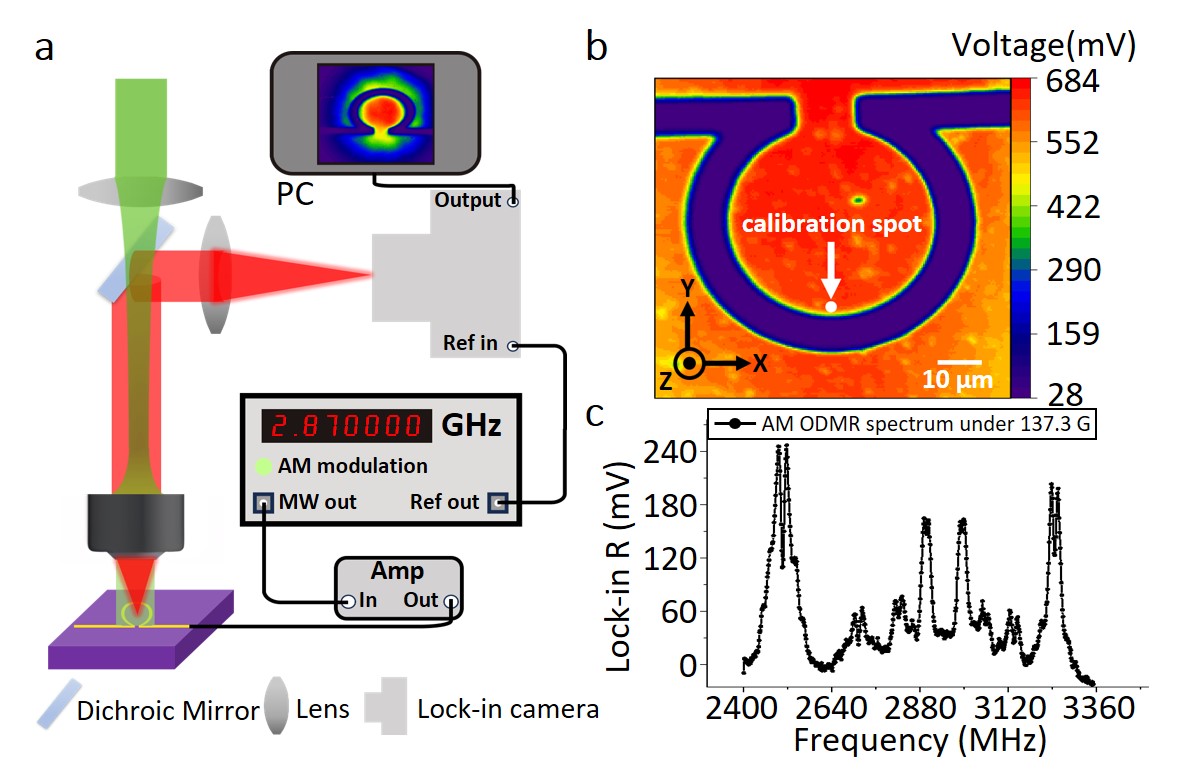}
    \caption{(a) Schematic diagram of the vector microwave field imaging setup. The optical system comprises a laser confocal scanning system and a wide-field imaging system. In the wide-field system, a lens is placed in front of the beam splitter to ensure uniform illumination of the spot on the sample in Köhler illumination mode, and the spot covers the ``$\Omega$"-shaped antenna area. (b) Confocal microscopy image. The dark fluorescent region corresponds to the antenna. The white dot indicates the interrogated region. The coordinate system is defined as indicated: the x- and y-axes lie parallel to the diamond surface, whereas the z-axis is oriented perpendicularly to surface. (c) After applying a biased magnetic field, the AM ODMR spectrum exhibit splittings at each peak under modulated MW irradiation.}
    \label{fig:Experimental_setup}
\end{figure}

As a proof of concept, we first calibrate the responsivity of MWs with our method in comparison to Rabi oscillations in a confocal setup, and then map the MW field distribution with a wide-field imaging system. The experimental schematic is shown in Fig.\,\ref{fig:Experimental_setup}\,a. The diamond is illuminated with a laser beam that is enlarged with a lens placed before the dichroic mirror and the objective to provide uniform, glare-free illumination, following the design of Köhler illumination design. The fluorescence is detected with a Lock-In camera that is capable of demodulation at each pixel. The diamond sample used for imaging is a dense sample with nitrogen concentration approximately 200\,ppm and NV center concentration approximately 15$\sim$37\,ppm\,\cite{li2021determination}. An ``$\Omega$''-shaped gold antenna, with a thickness of 50\,nm, an inner diameter of 80\,$\rm{\mu m}$, and a width of 12.5\,$\rm{\mu m}$, is deposited on the diamond surface to deliver MWs.
In the confocal system, a galvo mirror in the excitation light path is employed to walk the laser beam. The fluorescence is detected with a photodetector (PD) that converts the optical signal to voltage then sent to a LIA for demodulation. 

An image of the detected fluorescence is presented in Fig.\,\ref{fig:Experimental_setup}\,b. The dark region corresponds to the area covered by the $\Omega$ antenna that blocks the photons emitted from NVs directly under it. The laser spot was positioned at the white dot marked in the plot, and the laser power was set to 13\,$\rm{\mu W}$. An external magnetic field of 137.3\,G was applied to lift the degeneracy of NV's energy levels. An auxiliary square wave generated with a function generator was used as the AM source of the MW signal. The modulation rate was set to be 2.3\,kHz and the depth is 80\,\%. This modulated MW signal is amplified and transferred to the omega antenna. The laser power remained constant for the overall measurement. The fluorescence signal detected by the PD and demodulated with the LIA. Simultaneously, a replication of the auxiliary square wave was used as the demodulation reference signal in the LIA for demodulation. By stepping the MW frequency, a complete AM ODMR spectrum can be obtained. A typical AM ODMR for four-orientation NVs is presented in Fig.\,\ref{fig:Experimental_setup}\,c. In the presence of an external field, the AM ODMR spectrum displays eight groups of spectral features correspondingly.

\begin{figure}[H]
    \centering
    \includegraphics[width=1\linewidth]{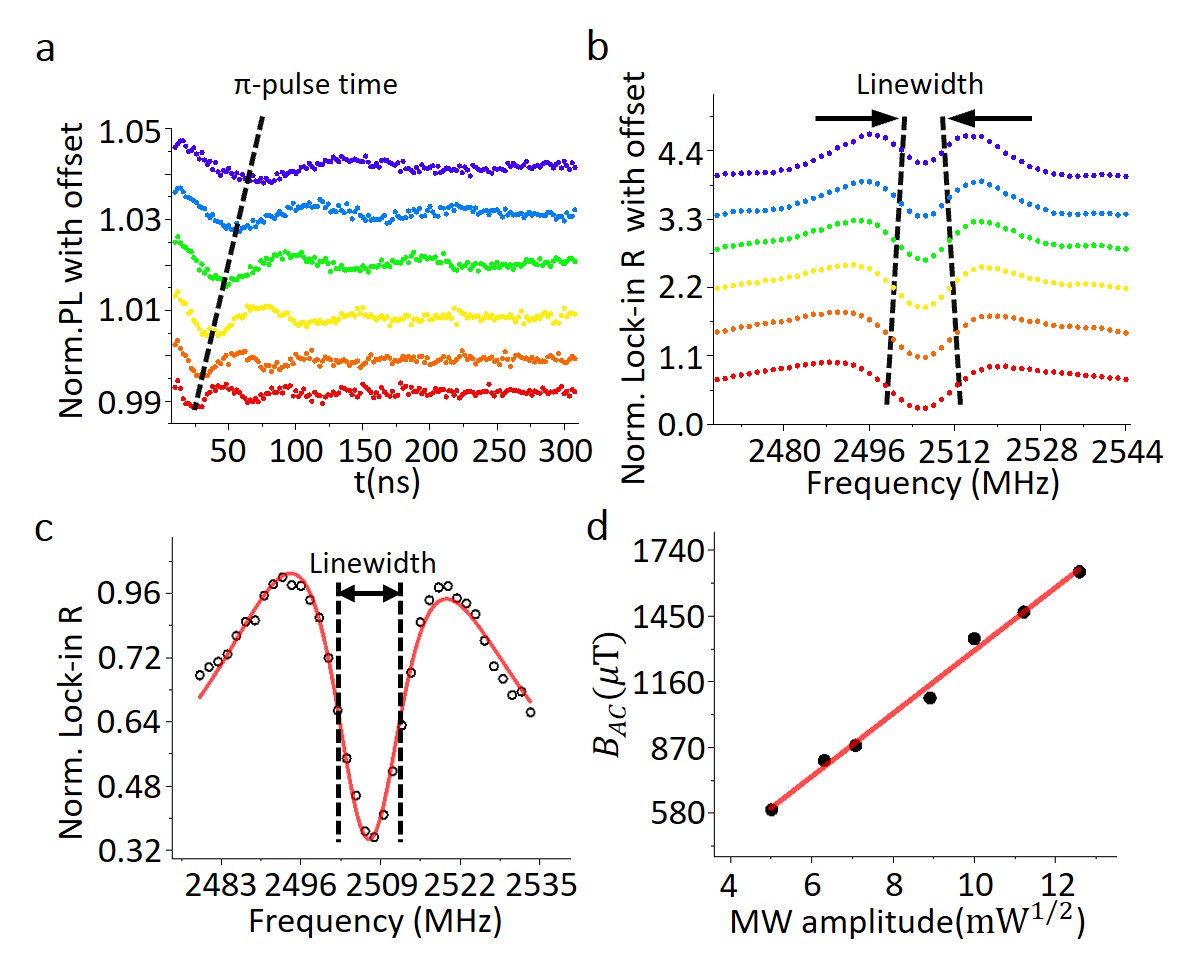}
    \caption{(a-b) The Rabi oscillations and AM ODMR spectral lines of NV centers with spatial orientations closest to the direction of the external magnetic field under varying MW power levels. The MW power input to the omega antenna, corresponding to the data lines from top to bottom, is 15.8\,mW, 25.1\,mW, 39.8\,mW, 63.1\,mW, 100\,mW, and 158.5\,mW, respectively. (c) The data and fitting curve of the left peak in the AM ODMR shown in Fig.\,\ref{fig:Experimental_setup}\,c. (d) The calibrated magnetic component of the MW field detected by NVs as a function of the applied MW powers. The data is fitted with a stringent linear function.}
    \label{fig:confocal}
\end{figure}

For calibration, we performed Rabi oscillations and AM ODMR measurements with the confocal system. Under a series of MW powers, the Rabi oscillation and AM ODMR data for NVs along four distinct crystallographic orientations are recorded. Figures\,\ref{fig:confocal}\,a and b present the experimental results of the NV centers that are oriented the most close to the direction of the external magnetic field. The results show that, similar to the Rabi oscillation frequency, the linewidth of AM ODMR spectral line increases linearly with the magnitude of MW, but with a larger dynamic range. 

The black dots in Fig.\,\ref{fig:confocal}\,c represent normalized experimental measurements of the left peak of the AM ODMR spectrum shown in Fig.\,\ref{fig:Experimental_setup}\,c. These data points were fitted using Eq.\,\eqref{eq3} (for detailed information, see the section\,\Rmnum{2} in the supplementary materials). The fitting results are presented as the red curve in Fig.\,\ref{fig:confocal}\,c, demonstrating an excellent agreement between the experimental results and the theoretical simulation. After the fitting parameters obtained, the linewidth is determined by taking the turning points of the second-order differential equations $\frac{\partial^{2}y}{\partial x^{2}}=0$ that correspond to the two points of the maximum slope. 
We also conducted the calibration for all the four-orientation NV centers (for details, see the section \Rmnum{3} in the supplementary materials). The results show that regardless of the orientations, the linewidths of the AM ODMR measured on all NVs exhibit linear dependence on the projected amplitude of applied MW field. 
The amplitude of the total vector MW field is calculated and presented in Fig.\,\ref{fig:confocal}\,d (for detailed information, see the section\,\Rmnum{1} in the supplementary materials). The data is fitted with a linear function, indicating a stringent linear response with our method. 

The sensitivity of the MW field measurement for this scheme can be estimated using the following formula\,\cite{schoenfeld2011real,lamba2024vector}:
\begin{equation}
\eta=\delta_{B}\cdot\sqrt{t}\label{eq3}
\end{equation}
where\,$\delta_{B}$ denotes the uncertainty in the amplitude measurement of the MW field, and \textit{t} represents the total experimental acquisition time per frequency point. The overall uncertainty in the amplitude of the MW is calculated according to the error propagation rules (see section\,\Rmnum{2} in the supplementary materials). The single-axis optimal MW field detection sensitivity achieved through evaluation is 600\,$\rm{n T/\sqrt{Hz}}$. 

Note that, apart from the main ODMR transitions, there may be additional sidebands under extreme high MW power, due to simultaneous spin transitions of NVs and other paramagnetic defects in the diamond, e.g. P1 centers\,\cite{simanovskaia2013sidebands}. It will reduce the dynamic range of our sensing regime. However, this can be mitigated by applying an auxiliary radio-frequency field to flip the spins of the defects and minimize the interactions to NVs. 



\section{Wide-field imaging of microwave}

After the validation of our sensing scheme, we performed vector mapping of MW fields with a wide-field-of-view imaging setup. 
The laser beam with a diameter of 300 \,$\rm{\mu}m$ uniformly illuminates the diamond sample and covers the entire antenna region. The spatial resolution of the imaging system is about of 800\,nm. The modulation is set to be 80\,\% in depth and 0.4 kHz for the modulating rate. 
The modulating square-wave is duplicated and sent to the Lock-In Camera as reference. Thus, the fluorescence can be demodulated at each pixel and a mapping of the MW field generated by the antenna can be achieved. 

Considering the differences in quantum efficiency between the Lock-In Camera and the PD in the confocal system, as well as the modulation rates used in the two experimental setups, it is essential to calibrate the responsivity of the wide-field experiment. The relationship between the MW field amplitude at the NV centers at the white spot (in Fig.\,\ref{fig:Experimental_setup}\,b) and the applied MW power is obtained via the comparison to that established in the confocal experiment. 

\begin{figure}
    \centering
    \includegraphics[width=1\linewidth]{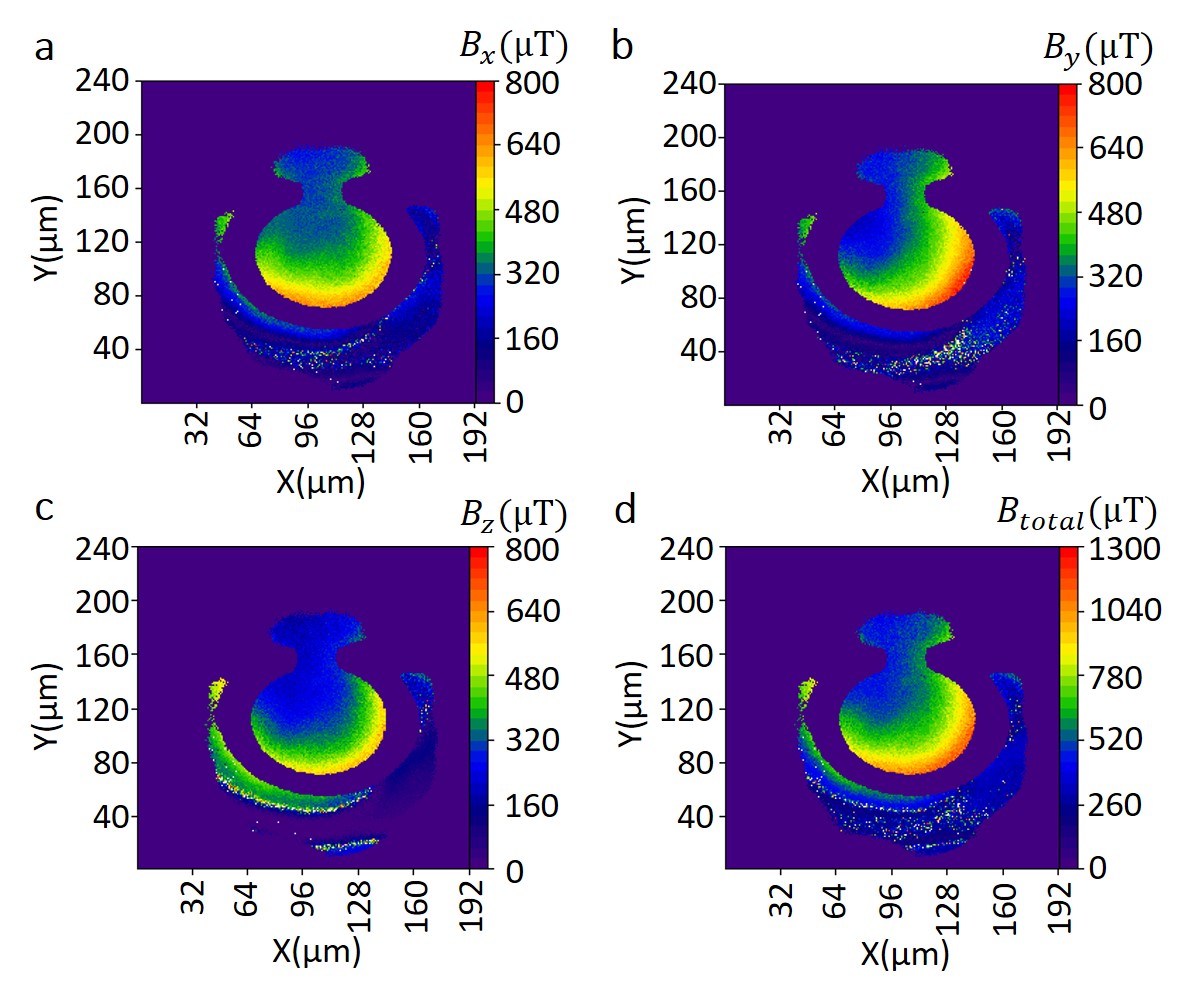}
    \caption{(a-d) The distribution of the microwave field generated by the omega antenna was examined in the x, y, and z directions, along with the spatial distribution of the total intensity.}
    \label{fig:widefield}
\end{figure}

Similarly to confocal measurement, the imaging setup is capable of measuring all the resonances of NVs in four orientations and is therefore extended for a vector mapping of MWs. By extracting the linewidth of the spectra at each pixel point in the lock-in camera, a full vector reconstruction of the MW fields delivered by the $\Omega$ antenna is achieved and mapped with a resolution of 800\,nm. 
The results are presented in Fig.\,\ref{fig:widefield}. Specifically, Fig.\,\ref{fig:widefield}\,a, b, and c correspond to the components in the x, y, and z directions (as defined in Fig.\,\ref{fig:Experimental_setup}\,b), respectively. 
The magnitude is shown in Fig.\,\ref{fig:widefield}\,d.

As shown in the results, the intensity of MW at the edge of the metallic wire is substantially higher than in the central region and the open gap of $\Omega$. Within an area of 80\,$\rm{\mu}m^2$, the amplitude of the MW field varies by 63\,\%, exhibiting significant spatial heterogeneity. 
The MW field components in the x- and y-directions display a pronounced spatial asymmetry, revealing a obvious imperfection in antenna fabrication. In the z-direction, the field intensity decreases rapidly from the edge toward the central region, indicative of a spatial profile primarily influenced by the near-field magnetic component.
The intensity of the MW field transmitted along the z axis is moderately lower than that in x and y for a thin layer antenna of 50\,nm in thickness.

The results demonstrate the capability of simultaneous high-spatial-resolution imaging of near-field MWs with our scheme. It provides a tool for the optimization of MW devices in terms of transmission efficiency, microscopic impedance matching, etc. Moreover, the spatial resolution can be further improved beyond the optical diffraction limit by replacing bulk diamond with nanodiamonds (for detailed information, see the section\,\Rmnum{5} in the supplementary materials). 


\section{Conclusion}
In summary, we have proposed and demonstrated a protocol for imaging of microwave vectors using NV centers in diamond. Our scheme is based on the spectral broadening effects of microwaves on the spin transition resonances of NV centers. By integrating the scheme with wide-field imaging setup, we successfully achieved simultaneous high-resolution mapping of the near-field distribution of the vector MW field generated by an ``$\Omega$'' antenna. Unlike conventional methods based on Rabi oscillation, our approach is insensitive to detunings between the MW frequency and the NV center’s resonant frequency during the measurement, thus significantly reducing the stringent requirements on the uniformity of magnetic field and the stability of temperature during the measurement. Benefiting from the simplicity of our scheme, measurements can be reproduced on larger and shallow-layer sensors, resulting in further improved resolution and sensitivity. Furthermore, the scheme operates effectively over a broad range of MW powers and maintains robustness even when it exceeds the contrast saturating threshold of NV's spin transitions. Moreover, the sensitivity and dynamic range can easily be tunable by adjusting the modulation depth in LIA detection or the magnitude of the auxiliary microwave for heterodyne measurements. The lower band of the detectable power is limited by the intrinsic ODMR linewidth. Improvement can be achieved by narrowing the linewidth of spin transitions, for example, upgrading the quantum property of NVs via high-pressure and high-temperature annealing\,\cite{tang2025optically}, or extending the coherence via sophisticated quantum control\,\cite{bauch2018ultralong}.

Our work paves the way for practical applications of fast and large-dynamic-range microwave microscopy, for example, visualizing the scatting paths of surface waves for advanced integrated circuit characterization, capturing the dynamic evolution of magnons for the design of spin-wave-based devices, scanning the dielectric constant distribution in human tissues in real-time for medical diagnostics, etc. The removal of spin controls will efficiently reduce the complexity of designing portable diamond microscopes.

\section{Acknowledgements}
We thank Dmitry Budker for informative discussions and helpful advice. We would like to express our gratitude to Guibin Lan, Yiheng Wang and Yuxin Ma for their assistance in fabricating the bowtie structure antenna. This work was supported by Beijing Natural Science Foundation (L233021), Innovation Program for Quantum Science and Technology (2023ZD0301100).

\bibliographystyle{apsrev4-2-2}
\bibliography{reference}

\end{document}


\title{Supplementary Material: Detuning-insensitive wide-field imaging of vector microwave fields with diamond
sensors}

\author{Xiu-Qi Chen}
\author{Rui-Zhi Zhang}
\affiliation{Beijing National Laboratory for Condensed Matter Physics, Institute of Physics, Chinese Academy of Sciences, Beijing 100190, China}
\affiliation{School of Physical Sciences, University of Chinese Academy of Sciences, Beijing 100049, China}

\author{Gang-Qin Liu}
\affiliation{Beijing National Laboratory for Condensed Matter Physics, Institute of Physics, Chinese Academy of Sciences, Beijing 100190, China}

\author{Huijie Zheng}
\email{hjzheng@iphy.ac.cn}
\affiliation{Beijing National Laboratory for Condensed Matter Physics, Institute of Physics, Chinese Academy of Sciences, Beijing 100190, China}

\date{\today}

\pacs{Valid PACS appear here}
\maketitle

\setcounter{equation}{0}
\renewcommand{\theequation}{S-\arabic{equation}}

\section{Vector microwave magnetic field reconstruction calculation method}

For a vector microwave electromagnetic field, the perceived MW amplitude by each axial NV center is given as follows:
\begin{equation}
B_{mw}^{i}=\begin{vmatrix}\overrightarrow{B_{mw}}\times\overrightarrow{e^{i}_{NV}}\end{vmatrix}\label{SI_eq1}
\end{equation}
Where \textit{i} = 1, 2, 3, 4 represent the four orientations of the NV centers: [111], [-111], [1-11], and [-1-11]. Considering that the crystal orientation of diamond is [100], We define the z-axis along the [100] direction, the x-axis along the [010] direction, and the y-axis along the [001] direction. Based on Eq.\,\eqref{SI_eq1}, we can determine the x, y, and z components of the vector MW field as follows\,\cite{wang2015high}:
\begin{equation}
\setlength\abovedisplayskip{3pt}
\setlength\belowdisplayskip{3pt}
\left\{\begin{matrix}B_{x}=K\sqrt{\frac{L}{MK}}
   \\B_{y}=\sqrt{\frac{KM}{L}}
   \\B_{z}=M\sqrt{\frac{L}{MK}}
  
\end{matrix}\right.\label{SI_eq2}
\end{equation}
Where
\begin{footnotesize}
\begin{equation}
\setlength\abovedisplayskip{3pt}
\setlength\belowdisplayskip{3pt}
\left\{\begin{matrix}K=\frac{3}{2}(-B_{mw}^{[111]2}+B_{mw}^{[-111]2}+B_{mw}^{[1-11]2}-B_{mw}^{[-1-11]2})
   \\L=\frac{3}{2}(-B_{mw}^{[111]2}+B_{mw}^{[-111]2}-B_{mw}^{[1-11]2}+B_{mw}^{[-1-11]2})
   \\M=\frac{3}{2}(-B_{mw}^{[111]2}-B_{mw}^{[-111]2}+B_{mw}^{[1-11]2}+B_{mw}^{[-1-11]2})
  
\end{matrix}\right.\label{SI_eq3}
\end{equation}
\end{footnotesize}
Therefore, by analyzing the linewidth of the AM ODMR spectra corresponding to these four orientations, the projections of the MW field amplitude along each orientation can be individually determined. Combining Eq.\,\eqref{SI_eq2} and \eqref{SI_eq3}, the complete vector information of the MW field can be reconstructed.

\section{Data fitting processing and sensitivity evaluation of microwave detection}

\begin{figure}[H]
    \centering
    \includegraphics[width=0.75\linewidth]{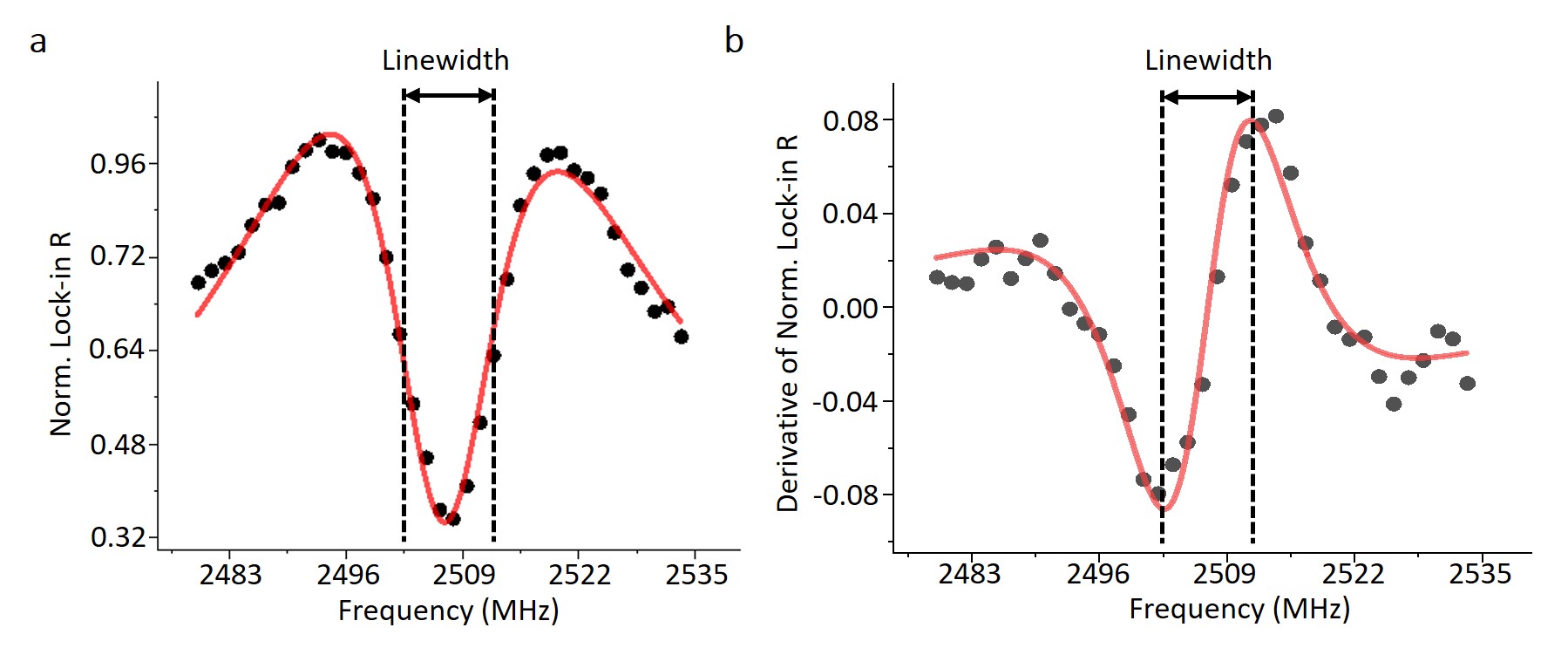}
    \caption{(a) The NV centers with spatial orientation most closely aligned to the external magnetic field direction manifested in the normalized AM ODMR spectrum under a MW modulation depth of 80\,\% , a modulation rate of 2.3\,kHz, and an integration time of 15\,ms per data point. (b) The derivative of the AM ODMR spectral line is used to determine its linewidth, which is defined as the distance between the two extreme points.}
    \label{fig:sensitivity_evaluation}
\end{figure}

We evaluated the microwave detection sensitivity of the proposed scheme by setting the output power of the MW signal to 100\,mW. An auxiliary source applied a 2.3\,kHz square wave modulation signal with a modulation depth of 80\,\%. The AM ODMR spectrum of the NV centers most closely aligned with the external magnetic field direction, obtained via stepping the MW frequency, is shown by the black dots in Fig.\,\ref{fig:sensitivity_evaluation}\,a. Each frequency point was measured with an integration time of 15\,ms. 
Based on the Lorentzian profile of CW-ODMR spectrum, the fitting function for the AM ODMR spectral line is formulated as follows:
\begin{footnotesize}
\begin{equation}
\setlength\abovedisplayskip{3pt}
\setlength\belowdisplayskip{3pt}
y=y_{0}+\frac{1}{\pi}\cdot\frac{2A_{1}\Delta\omega_{1}}{4\left(x-x_{c}\right)^{2}+\Delta\omega_{1}^{2}}-\frac{1}{\pi}\cdot\frac{2A_{2}\Delta\omega_{2}}{4\left(x-x_{c}\right)^{2}+\Delta\omega_{2}^{2}}\label{SI_eq4}
\end{equation}
\end{footnotesize}
The AM ODMR line is derived as the subtraction of two Lorentzian functions with different area ($A_{1}$ and $A_{2} $) and FWHM ($\Delta\omega_{1}$ and $\Delta\omega_{2} $) corresponding to the two different MW powers applied in the AM ODMR experiment. With a modulation depth of 80\,\%, according to Equation(2) in the main text, it follows $A_{1}\Delta\omega_{1}=5A_{2}\Delta\omega_{2}$.
We employed orthogonal regression to fit the measurement data, yielding the parameters\,$A_{1}$,\,$\Delta\omega_{1}$,\,$A_{2}$, and\,$\Delta\omega_{2}$ in Equation\,\eqref{SI_eq4} along with their corresponding uncertainties\,$e_{A_{1}}$,\,$e_{\Delta\omega_{1}}$,\,$e_{A_{2}}$, and\,$e_{\Delta\omega_{2}}$. The resulting fitted curve is displayed as the red line in Fig.\,\ref{fig:sensitivity_evaluation}\,a. Subsequently, we calculated the derivatives of both the experimentally collected data and the fitted curve, with the results shown in Fig.\,\ref{fig:sensitivity_evaluation}\,b. To optimize the MW detection sensitivity of the system, we adopted the distance between the two extreme points in Fig.\,\ref{fig:sensitivity_evaluation}\,b as a quantitative measure of the MW field amplitude. At the two extreme points, the following conditions must be satisfied:

\begin{footnotesize}
\begin{equation}
\setlength\abovedisplayskip{3pt}
\setlength\belowdisplayskip{3pt}
F(x,A_{1},\Delta\omega_{1},A_{2},\Delta\omega_{2})=\frac{16\,A_{2}\,\Delta\omega_{2}}{\pi \,{\left(4\,{\left(x-x_{c}\right)}^2+{\Delta\omega_{2}}^2\right)}^2}-\frac{16\,A_{1}\,\Delta\omega_{1}}{\pi \,{\left(4\,{\left(x-x_{c}\right)}^2+{\Delta\omega_{1}}^2\right)}^2}+\frac{4\,A_{1}\,\Delta\omega_{1}\,{\left(8\,x-8\,x_{c}\right)}^2}{\pi \,{\left(4\,{\left(x-x_{c}\right)}^2+{\Delta\omega_{1}}^2\right)}^3}-\frac{4\,A_{2}\,\Delta\omega_{2}\,{\left(8\,x-8\,x_{c}\right)}^2}{\pi \,{\left(4\,{\left(x-x_{c}\right)}^2+{\Delta\omega_{2}}^2\right)}^3}=0\label{SI_eq5}
\end{equation}
\end{footnotesize}
Based on Equation\,\eqref{SI_eq5}, the calculation can be performed as follows:
\begin{equation}
\setlength{\abovedisplayskip}{3pt}
\setlength{\belowdisplayskip}{3pt}
\left\{ \begin{aligned}
\frac{\partial x}{\partial A_{1}} &= -\frac{ \frac{\partial F}{\partial A_{1}} }{ \frac{\partial F}{\partial x} } \\
\frac{\partial x}{\partial \Delta\omega_{1}} &= -\frac{ \frac{\partial F}{\partial \Delta\omega_{1}} }{ \frac{\partial F}{\partial x} } \\
\frac{\partial x}{\partial A_{2}} &= -\frac{ \frac{\partial F}{\partial A_{2}} }{ \frac{\partial F}{\partial x} } \\
\frac{\partial x}{\partial \Delta\omega_{2}} &= -\frac{ \frac{\partial F}{\partial \Delta\omega_{2}} }{ \frac{\partial F}{\partial x} }
\end{aligned} \right. \label{SI_eq6}
\end{equation}
The error\,$e_{x}$ at the extreme point is as follows:
\begin{equation}
\setlength\abovedisplayskip{3pt}
\setlength\belowdisplayskip{3pt}
e_{x}=\sqrt{(\frac{\partial x}{\partial A_{1}})^{2}\cdot A_{1}^{2}+(\frac{\partial x}{\partial \Delta\omega_{1}})^{2}\cdot \Delta\omega_{1}^{2}+(\frac{\partial x}{\partial A_{2}})^{2}\cdot A_{2}^{2}+(\frac{\partial x}{\partial\Delta\omega_{2}})^{2}\cdot \Delta\omega_{2}^{2}}\label{12}
\end{equation}
Given that the linewidth of the AM ODMR spectrum is defined as the difference between the two extreme points\,$x_{1}$ and\,$x_{2}$, the fitting error of the linewidth can be expressed as: 
\begin{equation}
\setlength\abovedisplayskip{3pt}
\setlength\belowdisplayskip{3pt}
e_{linewidth}=\sqrt{(e_{x_{1}})^{2}+(e_{x_{2}})^{2}}\label{SI_eq7}
\end{equation}
The measurement error\,$\delta_{B}$ of the MW amplitude is determined by dividing \,$e_{linewidth}$ by the slope derived from the fitting in Fig.\,\ref{fig:S5_relation}\,c. Then, the MW measurement sensitivities for the four distinct NV center orientations can be obtained by \,$\eta=\delta_{B}\cdot\sqrt{t}$. When the MW signal source power is set to 100\,mW, the sensitivities of the four orientations of NV centers, evaluated using the aforementioned method, are measured to be 2.1, 0.6, 1.4, and 1.5\,$\rm{\mu T/\sqrt{Hz}}$, respectively. 

\section{The relationship between the linewidth of AM ODMR spectrum and the projected amplitude of microwave field of NV centers with four orientations}

\begin{figure}[H]
    \centering
    \includegraphics[width=1\linewidth]{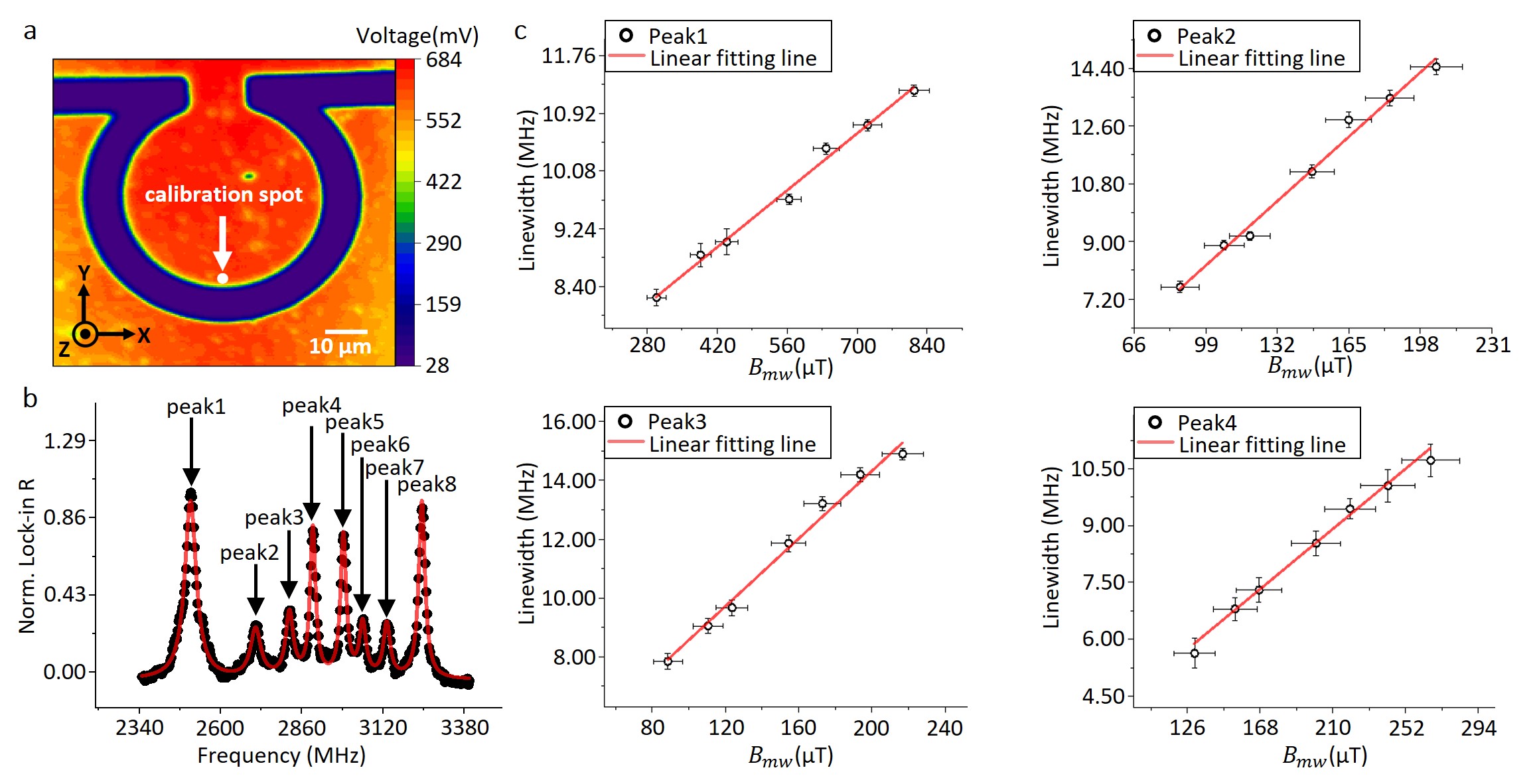}
    \caption{(a) Confocal microscopy image. (b) The normalized AM ODMR spectral lines under an external magnetic field of 137.3\,G, with MW modulation depth of 100\,\%. (c) At various MW power levels, the linewidth of the AM ODMR spectra of four different orientations of NV centers exhibits a strong linear correlation with the amplitude of the MW field, which is calibrated based on the Rabi oscillation frequency. }
    \label{fig:S5_relation}
\end{figure}

We positioned the laser spot of the home-built confocal system at the white dot 
 marked area in Fig.\,\ref{fig:S5_relation}\,a and applied an external magnetic field of 137.3\,G to resolve the energy levels corresponding to the four crystallographic orientations of NV centers. Fig.\,\ref{fig:S5_relation}\,b presents the AM ODMR spectral lines obtained under the application of an amplitude-modulated microwave with a modulation depth of 100\,\%. As an experimental verification of the vector microwave magnetometer principle we proposed, we first carried out Rabi oscillation experiments. We applied MW frequencies that were individually resonant with the energy levels of the nitrogen-vacancy (NV) centers in four distinct orientations. The MW power ranged from 15.8\,mW to 158.5\,mW. Due to the high nitrogen concentration in the S5 sample, its coherence time \,$T_{1\rho}$ is relatively short(\,$\sim$\,90\,ns). Therefore, lower MW power would lead to significant measurement errors. We carefully selected the Rabi oscillation frequencies measured under MW powers spanning from 39.8 mW to 158.5 mW for fitting. By utilizing the fitted Rabi oscillation frequencies, we were capable of calibrating the projections of the MW field along the axes of the NV centers in each of the four spatial orientations\,\cite{wang2015high}. This enabled us to establish the relationship between the projected amplitude of the MW field and the input power to the omega antenna. The input power to the omega antenna is expected to be proportional to the square of the MW field amplitude.

Subsequently, the MW power was increased from 15.8\,mW to 158.5\,mW, and amplitude-modulated MW signals with a modulation rate of 2.3\,kHz and a modulation depth of 80\,\% were applied. The AM ODMR measurement was then performed by stepping the MW frequency across the resonance range. Based on the AM ODMR spectral lines, the linewidth of the AM ODMR signals from the NV centers with four distinct spatial orientations can be extracted. Using the previously calibrated relationship between MW power and the corresponding projected amplitude for each orientation, the dependence of the linewidth of the AM ODMR spectrum on the MW field amplitude is obtained, as illustrated in Fig.\,\ref{fig:S5_relation}\,c. The experimental results show that for NV centers with four distinct orientations, the linewidth of the AM ODMR spectral lines all exhibits a good linear relationship with the corresponding projected amplitude of the MW field. This demonstrates the feasibility of employing this scheme for the measurement of the vector MW fields. However, further analysis reveals that the slopes of the four fitted lines are significantly different. Laser excitation, MW driving, magnetic field gradient and\,$T_{2}^{*}$\,dephasing time all contribute comparably to the linewidth of the ODMR spectrum\,\cite{2011Avoiding}. For the NV centers with four distinct spatial orientations, the applied bias magnetic field exhibits orientation-dependent magnetic field gradients. The presence of a magnetic field gradient results in broadening of the ODMR spectral lines and a corresponding reduction in \,$T_{2}^{*}$\,dephasing times\,\cite{barry2020sensitivity}. Additionally, variations in NV center concentrations among different orientations result in differing dipolar interactions\,\cite{kucsko2018critical}, which further influence\,$T_{2}^{*}$\,. These orientation-specific\,$T_{2}^{*}$\,dephasing times and magnetic field gradient give rise to intrinsic differences in the linewidth of the ODMR spectrum, thereby causing the slope of the linewidth of AM ODMR to vary with increasing MW amplitude. Although these factors may lead to different variations in the linewidths of the AM ODMR spectra across NV centers with different orientations as the MW field amplitude changes, they do not compromise the linearity of the response and therefore do not affect the accuracy of the measurement scheme.


\section{Simulation results of the relationship between microwave field amplitude and linewidth of AM ODMR spectra under different microwave modulation depths}

\begin{figure}[H]
    \centering
    \includegraphics[width=0.5\linewidth]{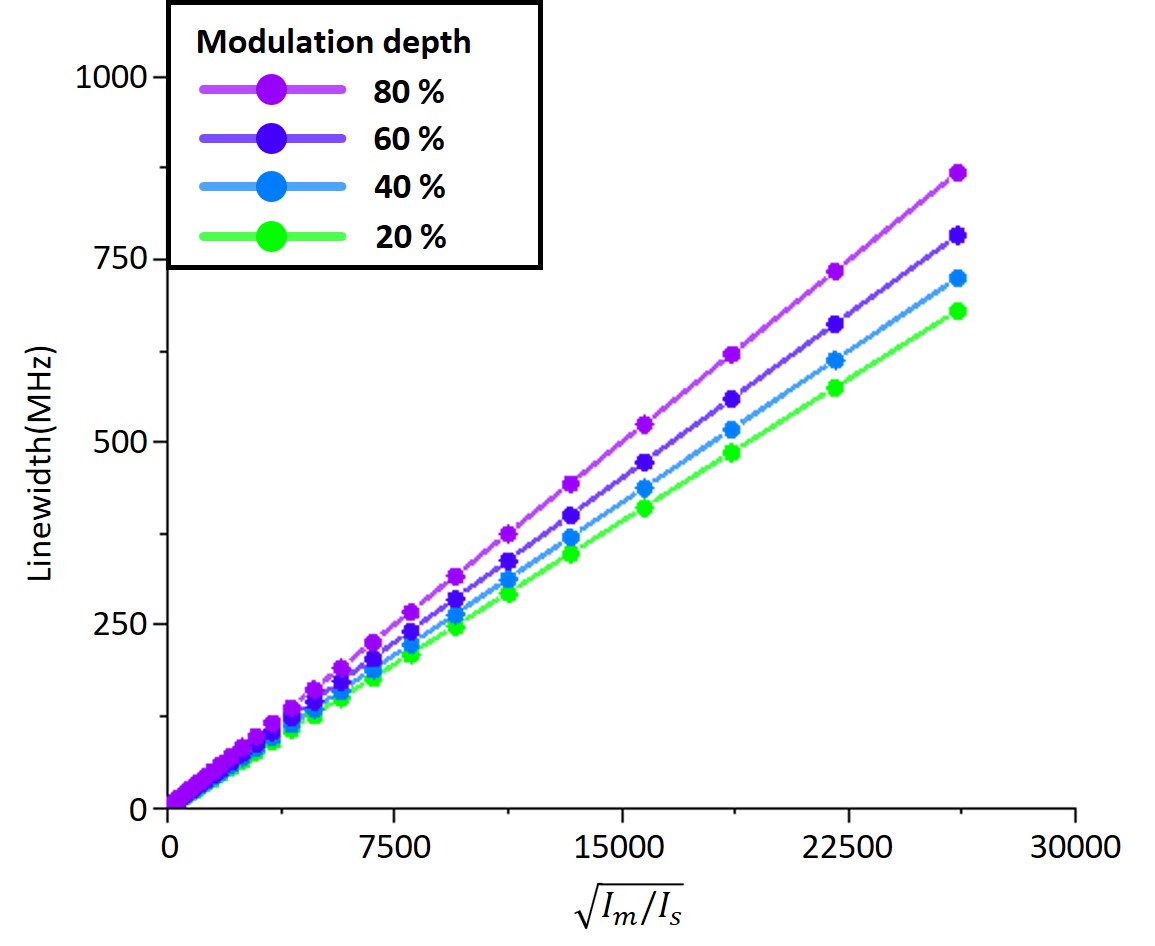}
    \caption{The relationship between the linewidth of AM ODMR spectral lines and the amplitude of the microwave field varies with different microwave modulation depths.}
    \label{fig:change modulation depth}
\end{figure}

Fig.\,\ref{fig:change modulation depth} presents the numerical simulation results of the relationship between the linewidth of AM ODMR spectral lines and the relative microwave field amplitude under different modulation depths. The results demonstrate that, across various modulation depths, the spectral linewidth all exhibits a strong linear dependence on the MW field amplitude. Moreover, as the modulation depth decreases, a given change in MW field amplitude produces a smaller change in the spectral linewidth. This suggests that for applications demanding higher-intensity microwave fields, decreasing the modulation depth can enhance measurement performance. Furthermore, since $I_{S}=\Gamma^{\infty}_{p}\Gamma^{\infty}_{c}\left(\frac{S}{1+S}\right)^{2}$, increasing the laser power—and thus the saturation parameter $S$ of the radiative transition—can enhance the detection capability for stronger microwave signals. While meeting the demand for higher sensitivity, the laser power should be reduced to make\,$I_{S}$ smaller, thereby obtaining a greater response of linewidth to magnitude.

\section{The implementation of the scheme on nanodiamonds}

\begin{figure}[H]
    \centering
    \includegraphics[width=1\linewidth]{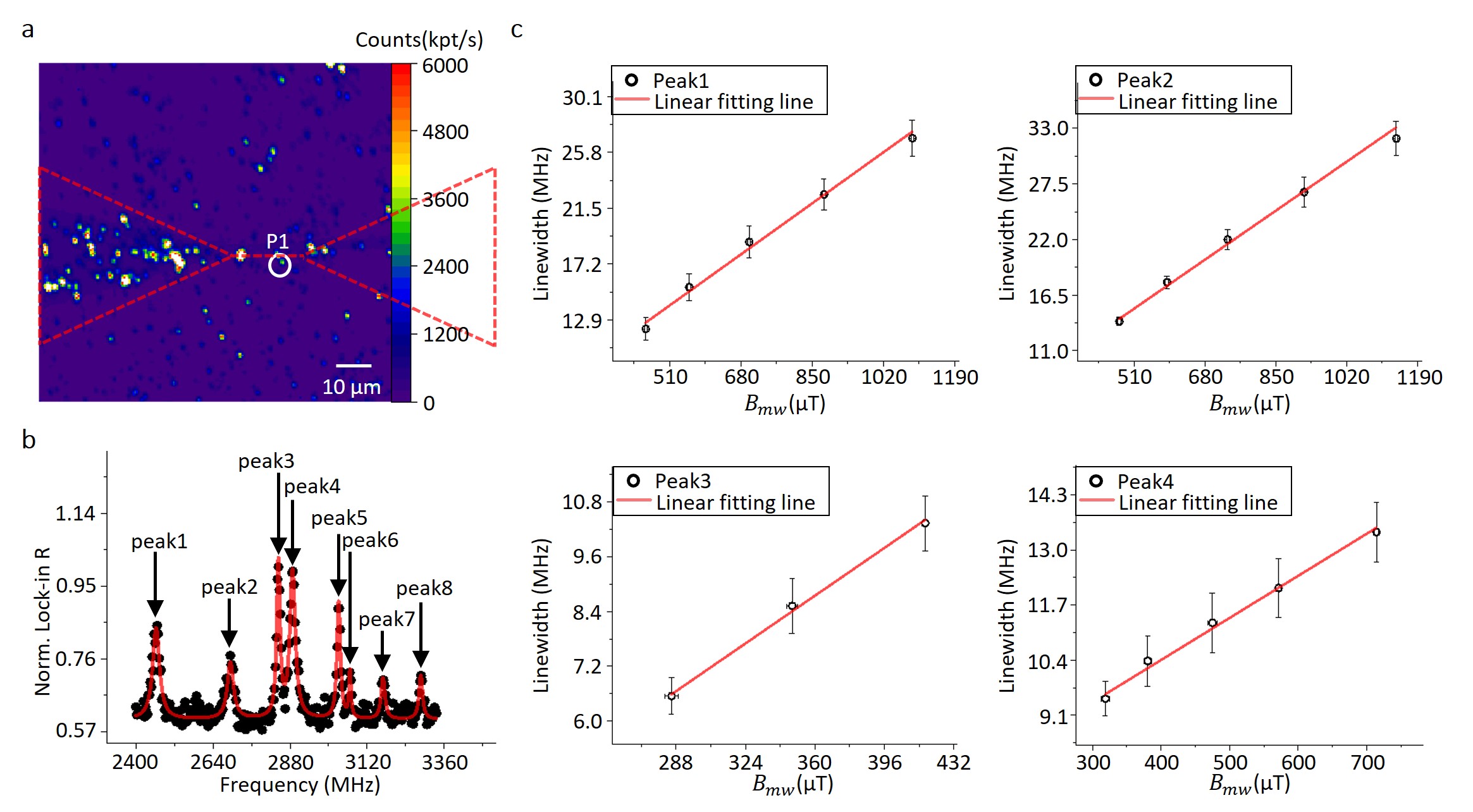}
    \caption{(a) Confocal microscopy image. The diamond used in the experiment is indicated by a white circle in the figure and labeled as P1. (b) The normalized AM ODMR spectral lines of P1 under an external magnetic field of 152.4\,G, with MW modulation depth of 100\,\%. (c) The linear relationship between the linewidth of the AM ODMR spectral lines of NV centers in the four orientations of P1 diamond and the amplitude of the MW magnetic field. }
    \label{fig:ND_relation}
\end{figure}
To further demonstrate the scheme's potential for achieving high spatial resolution, we conducted experiments using nanodiamonds($\left[NV\right]\sim3\,\rm{ppm}$) with an average size of 100\,nm. A gold nanowire-bowtie structure(marked by the red dotted line in the Fig.\,\ref{fig:ND_relation}\,a) was fabricated on a silicon wafer via magnetron sputtering to enhance MW confinement\,\cite{chen2024microwave}. Subsequently, a nanodiamond solution (concentration\,$\sim5\,\rm{\mu g/ml}$) was deposited onto the substrate. The resulting fluorescence scanning image is presented in Fig.\,\ref{fig:ND_relation}\,a, showing uniform dispersion of the nanodiamonds. The region enclosed by the dashed line corresponds to the location of the nanowire-bowtie structure. The specific nanodiamond used in our measurements is labeled as P1 in the figure. Under an applied bias magnetic field of 152.4 G, the normalized AM ODMR spectrum with a modulation depth of 100\,\% is shown in Fig.\,\ref{fig:ND_relation}\,b. After the degeneracy of the energy levels associated with the four orientations of NV centers was lifted by an external magnetic field, we systematically varied the MW source power to 251.2, 398.1, 631.0, 1000.0, and 1584.9\,mW, respectively, and performed AM ODMR measurements using a modulation depth of 50\,\% and a modulation rate of 1.3\,kHz. At fixed MW power level, Rabi oscillation experiments were separately conducted on NV centers with four distinct orientations to determine the corresponding projected MW amplitudes for each orientation. Fig.\,\ref{fig:ND_relation}\,c illustrates the relationship between the linewidth of the AM ODMR spectra for the four NV center orientations and the corresponding projected amplitude of the MW field along each spatial orientation. A clear linear correlation is observed across all orientations. This suggests that our approach facilitates the utilization of nanodiamonds to substantially improve the spatial resolution of vector microwave field detection, with the potential to achieve sub-100-nanometer resolution or higher.

\bibliographystyle{apsrev4-2-2}
\bibliography{reference}